\documentclass[aip,pof,reprint]{revtex4-1}
\usepackage{graphicx}

\usepackage{bm}
\usepackage{amsmath}
\usepackage{amssymb}
\usepackage{amsfonts}
\usepackage{amsbsy}

\begin{document}

\title{Conditional Eulerian and Lagrangian velocity increment
  statistics of fully developed turbulent flow} 

\author{Holger 
   \surname{Homann}$^{1,2}$ , Daniel \surname{Schulz}$^1$, and Rainer
  \surname{Grauer}$^1$}  
\email{holger@tp1.rub.de}
\affiliation{$^1$
  Theoretische Physik I, Ruhr-Universit\"at, 44780 Bochum,
  Germany\\ $^2$ Universit\'e de Nice-Sophia Antipolis, CNRS,
  Observatoire de la C\^ote d'Azur, Laboratoire Cassiop\'ee, Bd.\ de
  l'Observatoire, 06300 Nice, France}

\date{\today}

\begin{abstract}
Conditional statistics of homogeneous isotropic turbulent flow is
investigated by means of high-Reynolds number direct numerical
simulations performed with $2048^3$ collocation points. Eulerian as
well as Lagrangian velocity increment statistics under several
conditions are analyzed and compared. In agreement with experimental
data longitudinal probability density functions $P(\delta^{||}_l
u|\epsilon_l)$ conditioned on a scale-averaged energy dissipation rate
are close to Gaussian distributions over all scales within the
inertial range of scales. Also transverse increments conditioned on
either the dissipation rate or the square of the vorticity have
quasi-Gaussian probability distribution functions (PDFs). Concerning
Lagrangian statistics we found that conditioning on a trajectory
averaged energy-dissipation rate $\epsilon_\tau$ significantly reduces
the scale dependence of the increment PDFs $P(\delta_\tau
u_i|\epsilon_\tau)$. By means of dimensional arguments we propose a
novel condition for Lagrangian increments which is shown to reduce
even more the flatness of the corresponding PDFs and thus
intermittency in the inertial range of scales. The conditioned
Lagrangian PDF corresponding to the smallest increment considered is
reasonably well described by the K41-prediction of the PDF of
acceleration. Conditioned structure functions show approximately
K41-scaling with a larger scaling range than the unconditioned ones.

\end{abstract}



\maketitle

\noindent
\section{Introduction}
The problem of anomalous scaling can be seen as one of the great
unsolved problems in turbulence research. The scaling laws of velocity
structure functions $S_p(l) = \langle (\delta_l u)^p\rangle \sim
l^{\zeta_p}$ within the inertial range of scales of fully developed
turbulence has inspired a lot of publications over the last decades
\cite{frisch:1995,tsinober:2009}. The velocity increment under
consideration is usually either the longitudinal $\delta^{||}_l
u=(\bm{u}(\bm{x}+\bm{l})-\bm{u}(\bm{x}))\cdot \bm{\hat{l}}$ or the
transverse one $\delta^{\perp}_l
u=|(\bm{u}(\bm{x}+\bm{l})-\bm{u}(\bm{x}))\times \bm{\hat{l}}|$. Both
are so called Eulerian increments because the velocity differences are
taken over spatial separations at a the same instant of
time. Kolmogorov's K41-theory\cite{K41} implies a linear scaling law
$\zeta_p=p/3$ not distinguishing between the two different types of
increments mentioned before. However, direct numerical simulations
(DNS) and experiments show a deviation of the form\cite{frisch:1995}
\begin{equation}
  \label{eq:correction}
  \zeta_p=p/3-\mu_p,
\end{equation}
with a positive $\mu_p$. The question whether longitudinal and
transverse statistics possesses two different sets of scaling
exponents $\zeta^{||}_p$, $\zeta^{\perp}_p$ respectively, is still
under discussion.  Experimental
observations~\cite{Dhruva2000,Zhou2005a} as well as
DNS~\cite{Grossmann1997} found slightly smaller scaling exponents for
the high-order transverse than for the longitudinal structure
functions. It is not yet clear whether these findings are
finite-Reynolds number and/or anisotropy effects. In the case of
election-MHD turbulence~\cite{Germaschewski1999} the differences
between longitudinal and transverse scaling exponents were found to
decrease with Reynolds-number. One has also to be very careful in the
determination of these scaling
exponents~\cite{grauer-homann-etal:2010} as longitudinal and
transverse structure functions possess differing scaling ranges.

Guided by Kolmogorov's refined self similarity hypothesis\cite{K62}
(RSH)
\begin{equation}
  \label{eq:rsh}
  \delta^{||}_lu_=\beta_1(\epsilon_l l)^{\frac{1}{3}}
\end{equation}
which states a relation between the local energy dissipation rate
\begin{equation}
  \label{eq:localEnergyDiss}
  \epsilon(\bm{x}) = \nu \sum_{i,j}[\partial_j
    u_i(\bm{x})+\partial_i u_j(\bm{x})]^2
\end{equation}
averaged over a scale $l$
\begin{equation}
  \epsilon_l=\frac{1}{l}
  \int^l_0\epsilon(\bm{x}+s\,\hat{\bm{l}}) ds
  \label{eqn:dissipation}
 \end{equation}
($l$ indicating the same line appearing in $\delta^{||}_l u$ and
$\delta^{\perp}_l u$) and the velocity fluctuation $\delta_l^{||}u$
over that scale Gagne et Ac.\cite{GAG94} experimentally measured
conditional velocity increment statistics. They found the probability
density functions (PDFs) $P(\delta_l^{||} u |\epsilon_l)$ to be nearly
Gaussian from the dissipation- up to the integral-scale which implies
a linear scaling law with $\mu_p = 0$ in (\ref{eq:correction}). It is
believed that anomalous scaling ($\mu_p \neq 0$) in turbulent flows has
its origin in small-scale intermittency of the local energy
dissipation rate. This point of view is supported by this result,
namely that the statistics of increments become Gaussian once they are
conditioned on a scale-averaged energy dissipation rate.

Recently, new experimental techniques have provoked a renewed interest
in Lagrangian
statistics~\cite{ott-mann:2000,porta-bodenschatz-etal:2001,mordant:2001}. Here,
velocity increments are taken along trajectories of fluid elements
(tracers). Lagrangian velocity increments are defined by
\begin{align}
  \begin{aligned}
    \label{eq:lagInc}
    \delta_\tau v_i &= v_i(\tau) - v_i(0) \\
    &= u_i(\bm{X}(\bm{x}_o,\tau),\tau)-u_i(\bm{X}(\bm{x}_0,0),0),
  \end{aligned}
\end{align}
where $\bm{X}(\bm{x}_0,\tau)$ denotes the trajectory of a tracer which
started at the position $\bm{x}_0$ at time $t=0$. Although one might
expect a scaling law of the corresponding structure functions
$S_p^L(l) = \langle (\delta_\tau v_i)^p\rangle \sim \tau^{\zeta^L_p}$
within the temporal inertial range of scales it has not yet been
clearly observed\cite{yeung-pope-etal:2006,Biferale2008}. The origin
of the strong measured intermittency, the very existence of an
inertial range of scales, the corresponding scaling exponents, and
their relation to Eulerian intermittency are still open
issues\cite{biferale:2004b,kamps-friedrich-grauer:2009, homann-kamps-etal:2009}. 

In this paper we measure conditional velocity statistics both in the
Eulerian as well as in the Lagrangian frame of reference by means of
high-Reynolds number DNS. We approve the experimental results obtained
by Gagne et al.~\cite{GAG94} and Naert et al.~\cite{NAE98} and
complement their findings by a detailed scale by scale analysis and an
investigation of the statistics of conditioned transverse velocity
increments. Furthermore we analyze conditioned Lagrangian increment
statistics.

In the next section we briefly present the numerical
method. Section~\ref{sec:euler} presents the results in the
Eulerian frame and section~\ref{sec:LAGRANGIAN} those in the
Lagrangian frame of reference. Conclusion are summarized in
section \ref{sec:CONLUSION}.

\section{Numerics}

\begingroup
\squeezetable
\begin{table*}
  \centering
  \begin{ruledtabular}
  \begin{tabular}{cccccccccccccc}
    $\Re_{\lambda}$&$u_\mathrm{rms}$& $\epsilon_\mathrm{k}$&$\nu$            & $dx$              & $\eta$           &$\tau_\eta$&$L$   &$T_L$& $N^3$     & $N_p$\\
    \hline 
    $460$         &$0.189$        &$3.6\cdot 10^{-3}$   &$2.5\cdot 10^{-5}$&$3.07\cdot 10^{-3}$&$1.45\cdot 10^{-3}$& 0.083     &1.85 & 9.9  &$2048^3$ & $10^{7}$
 \end{tabular}
  \end{ruledtabular}
 \caption{\label{table} Parameters of the numerical simulations.
    $\Re_\lambda = \sqrt{15VL/ \nu}$: Taylor-Reynolds number,
    $u_\mathrm{rms}$: root-mean-square velocity, $\epsilon_\mathrm{k}$: mean
    kinetic energy dissipation rate, $\nu$: kinematic viscosity,
    $dx$: grid-spacing, $\eta =(\nu^3/\epsilon_\mathrm{k})^{1/4}$: Kolmogorov dissipation length
    scale, $\tau_\eta = (\nu/\epsilon_\mathrm{k})^{1/2}$: Kolmogorov
    time scale, $L = (2/3E)^{3/2}/\epsilon_\mathrm{k}$: integral
    scale, $T_L = L/u_\mathrm{rms}$: large-eddy turnover time, $N^3$:
    number of collocation points, $N_p$: number of tracer particles.}
\end{table*}
\endgroup

The numerical simulations were performed by solving the incompressible
Navier-Stokes equations
\begin{align}
  \label{impuls}
  &\partial_t \bm{u} + (\bm{u} \cdot\nabla)\bm{u} = \bm{f} -\nabla p
  +\nu \Delta \bm{u} \\ &\nabla \cdot \bm{u} = 0,
\end{align}
in a periodic cube with a pseudo-spectral method using a high-order
exponential cut-off~\cite{hou-li:2007,grafke-homann-etal:2007}.

We parallelize the computations via a pencil geometry by means of the
San Diego P3D-FFT~\cite{p3dfft} and explore the
BlueGene/P-architecture (the $2048^3$ simulation was performed on 32k
processors). The time integration of the velocity field is done by
means of a strongly stable Runge-Kutta third order scheme
\cite{shu-osher:1988}. In order to maintain a statistically stationary
flow a forcing $\bm{f}$ is applied which keeps constant the modes of
the two lowest Fourier-shells. Averages are taken over several
statistically independent realizations of the velocity field.

Once a stationary state has been reached 10 Million tracers are seeded
into the flow and integrated according to
\begin{equation}
  \dot{\bm{X}}(\bm{x}_0,t)=\bm{u}(\bm{X}(\bm{x}_0,t),t),
\end{equation}
where $\bm{u}(\bm{X},t)$ is the velocity obtained from (\ref{impuls}). 

In order to obtain the velocity at the particle position from the grid
values we use a tri-cubic
interpolation~\cite{homann-dreher-etal:2007}. All relevant quantities
such as the gradient of velocity are stored in intervals of 1/7th of
the dissipation timescale. The main parameters of all simulations are
given in Table~\ref{table}.

\section{Eulerian conditional statistics}
In this section we present the result on conditioned Eulerian
increments statistics. We are going to start with longitudinal
increments, followed by transverse increments in the subsequent
section.
\label{sec:euler}
\subsection{Longitudinal increments}
\label{sec:LONGITUDINAL}
\begin{figure}[h]
  \begin{center}
    \includegraphics[width=0.8\columnwidth]{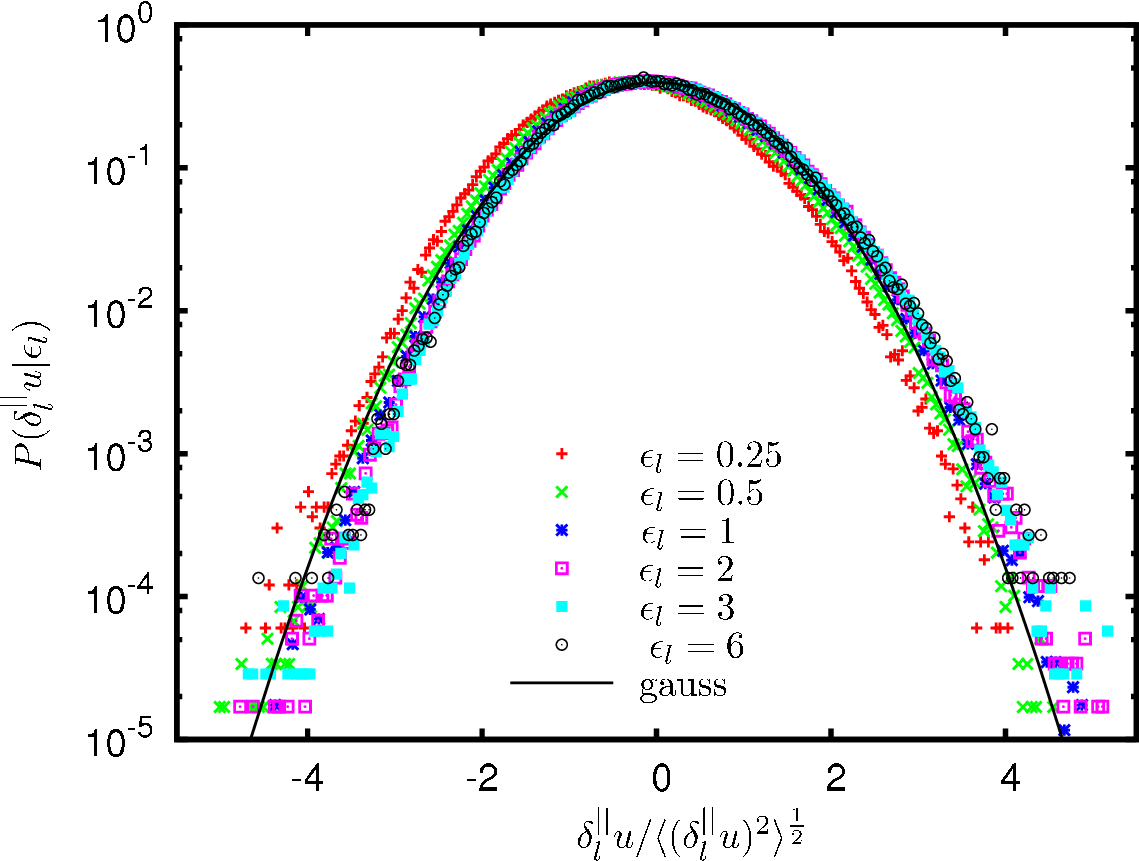} 
  \end{center}
  \caption{\label{fig:pdfs_epsilon} Conditioned PDFs $P(\delta^{||}_l
    u |\epsilon_l)$ for different space-averaged dissipation rates
    $\epsilon_l$ for $l=93\eta$ in comparison to a Gaussian
    distribution, $\epsilon_l=1$ corresponds to the most probable
    energy dissipation rate, the others of multiples of this
    rate. All PDFs are normalized to unit variance.}
\end{figure}

\begin{figure}[h]
  \begin{center}
    \includegraphics[width=0.8\columnwidth]{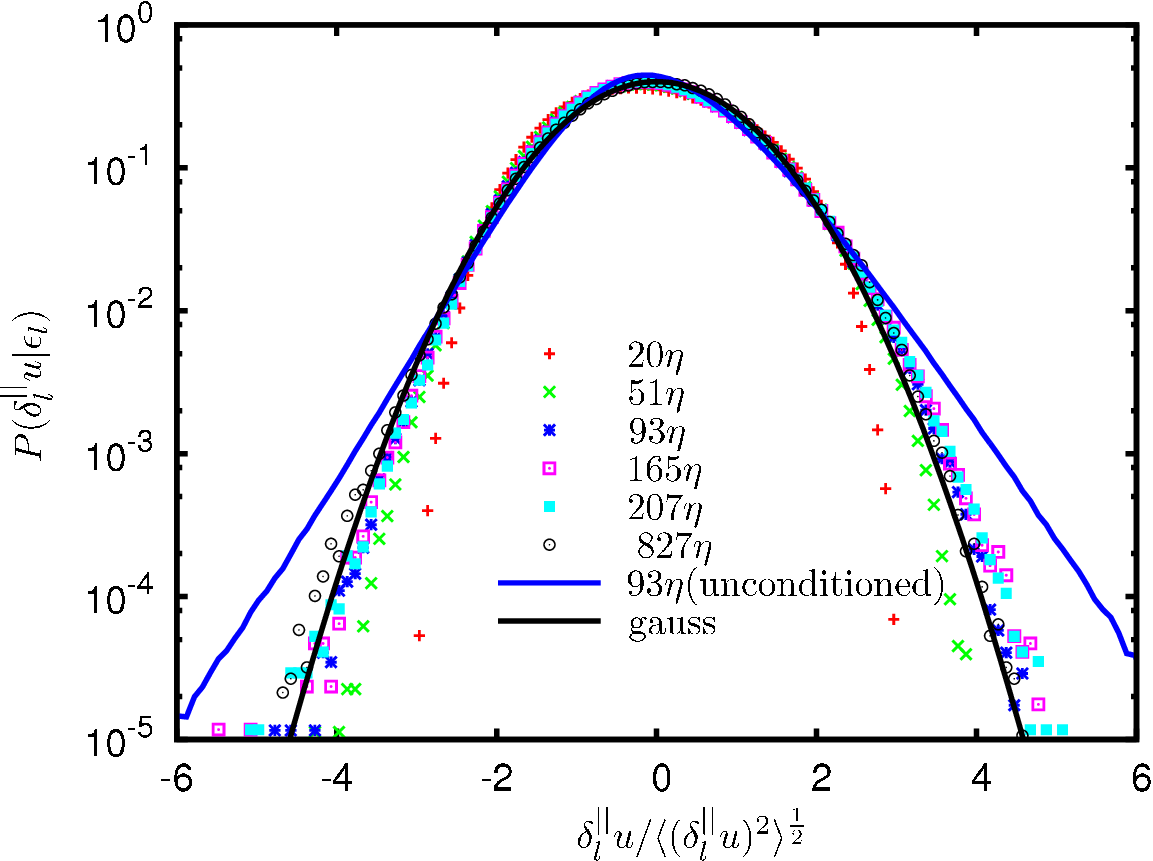} 
  \end{center}
  \caption{\label{fig:pdfs} Conditioned PDFs $P(\delta^{||}_l u
    |\epsilon_l)$ for different separations $l$ and the most probable
    value of $\epsilon$, normalized to unit variance}
\end{figure}

Following the experimental results obtained by Gagne et
al.~\cite{GAG94} as well as Naert et al.~\cite{NAE98}, we split our
simulation domain in subsets $\Omega_{\epsilon_l}$ of fixed rate of
energy dissipation $\epsilon_l$ on a line $l$ defined by
(\ref{eqn:dissipation}). On these subsets we consider longitudinal
velocity increments $\delta_l^{||} u$ in order to obtain conditional
PDFs $P(\delta_l^{||} u |\epsilon_l)$.  The standard (unconditioned)
PDFs can be recovered by integrating this PDF over all
$\epsilon_l$. In agreement with the experimental results we find for
a separation $l$ within the inertial range of scales nearly Gaussian
statistics for different dissipation rates (see
Fig.~\ref{fig:pdfs_epsilon}).
Also for different scales $l$ we recover Gaussianity (see
Fig.~\ref{fig:pdfs}). The unconditioned PDFs have clearly flatter
tails.

\begin{figure}[h]
  \begin{center}
    \includegraphics[width=0.8\columnwidth]{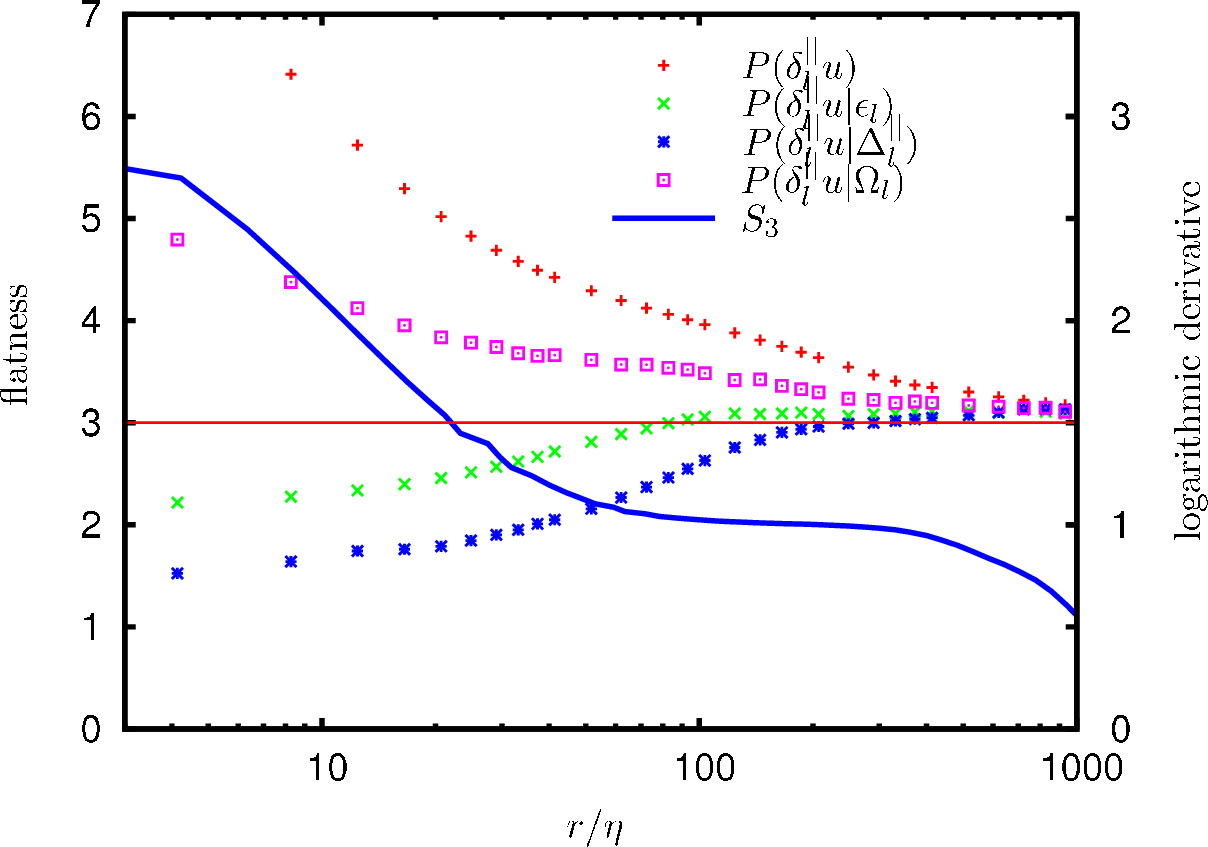} 
  \end{center}
  \caption{\label{fig:AllFlat} Flatness factors for conditioned
    velocity increment PDFs $P(\delta^{||}_l u |\epsilon_l)$,
    $P(\delta^{||}_l u |\Omega_l)$, $P(\delta^{||}_l u |\Delta_l)$ and
    the unconditioned PDF $P(\delta^{||}_l u)$, including the
    logarithmic derivative of the third order structure function
    $S^{||}_3$. The horizontal line indicates the flatness of a
    Gaussian distribution.}
\end{figure}

As a measure of the deviation from Gaussianity we present in
Fig.~\ref{fig:AllFlat} the flatness $\langle (\delta^{||}_l
u)^4\rangle/\langle (\delta^{||}_l u)^2\rangle$. The figure includes
the logarithmic derivative of the third-order structure
function $S_3^{||}(l)$ in order to illustrate the inertial range via
its plateau. The PDFs of the conditioned increments
(Fig.~\ref{fig:pdfs}) reach a flatness of approximately three
throughout the inertial range of scales.

For comparison we conditioned the velocity increments on other
quantities composed of velocity-gradient tensor elements, namely the
vorticity $\bm{\omega}=\nabla\times\bm{u}$ and the longitudinal
gradient $\hat{\bm{l}}\cdot\nabla \bm{u} $. As for the
energy dissipation rate we consider spatial averages of the square of
vorticity
\begin{equation}
  \Omega_l=\frac{1}{l} \int^l_0ds\, \nu |\bm{\omega}(\bm{x}+s\,\hat{\bm{l}})|^2.
  \label{equ:vorticityInt}
\end{equation}
and the square of the longitudinal gradient
\begin{equation}
  \Delta^{||}_l=\frac{1}{l} \int^l_0ds\, \nu |\hat{\bm{l}}\cdot\nabla
  \bm{u}(\bm{x}+s\,\hat{\bm{l}})|^2.
  \label{equ:gradientInt}
\end{equation}

From Fig.~\ref{fig:AllFlat} one recognizes that the flatness of
$P(\delta^{||}_l u |\epsilon_l)$ is closer to a Gaussian distribution
over all scales than $P(\delta^{||}_l u |\Omega_l)$ or
$P(\delta^{||}_l u |\Delta_l)$. It is interesting to remark that the
integral of the longitudinal gradient over $l$ is the longitudinal
increment. That the energy dissipation rate $\epsilon_l$ nevertheless
works better than this longitudinal gradient implies that correlations
of the form $\partial_j u_i \partial_i u_j$ with $i\neq j $ are
essential in the condition (\ref{eq:localEnergyDiss}) and in the RSH
(\ref{eq:rsh}).

\begin{figure}[h]
  \begin{center}
    \includegraphics[width=0.8\columnwidth]{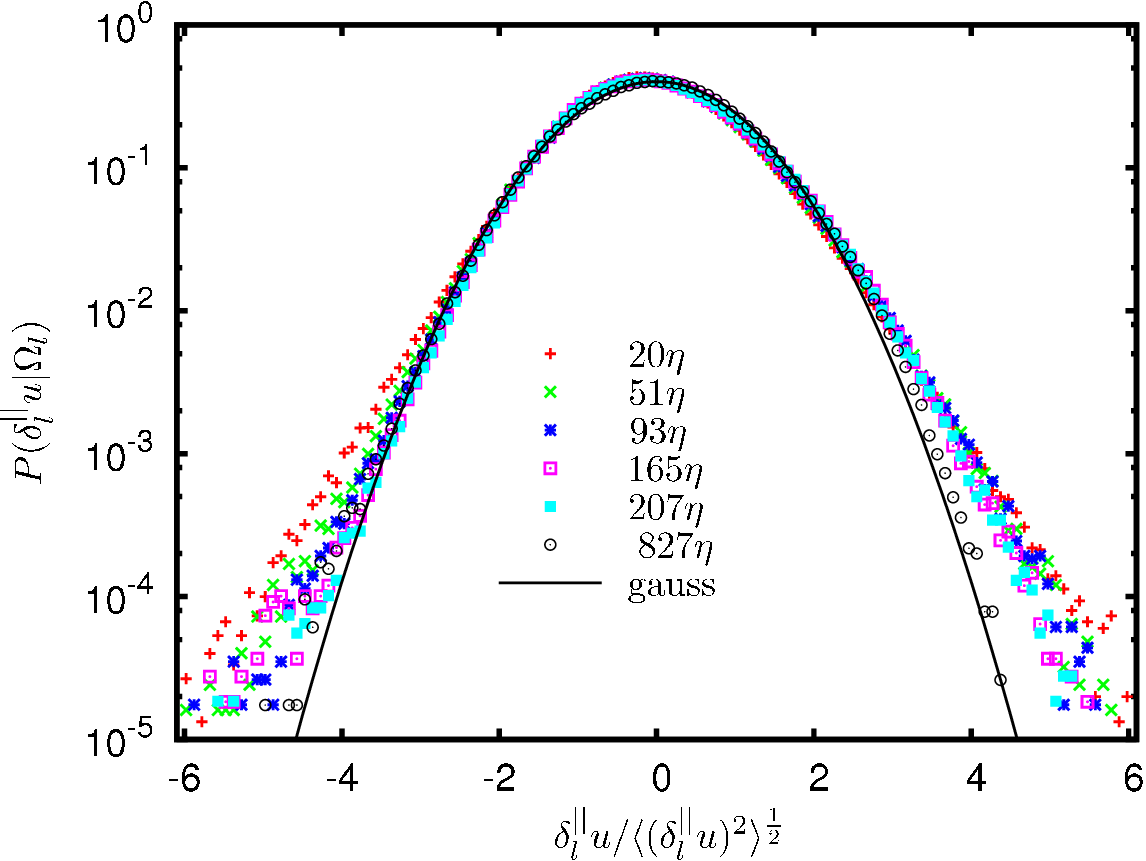} 
  \end{center}
  \caption{\label{fig:fig1_vort} Conditioned PDFs $P(\delta^{||}_l u
    |\Omega_l)$ for different separations $l$ in comparison to a
    Gaussian distribution, normalized to unit variance}
\end{figure}

The flatness of the PDFs conditioned to the scale-averaged energy
dissipation rate come closest to the Gaussian value. However, from
Fig.~\ref{fig:fig1_vort} one observes that the PDFs conditioned on
$\Omega_l$ are also nearly scale-invariant but not exactly Gaussian.

Whether longitudinal increments are conditioned on the
energy-dissipation or vorticity yields quasi-identical results. With
respect to the slightly smaller flatness of the former one can
conclude that longitudinal increment statistics is coupled more
closely to the scale averaged energy dissipation rate than to the
vorticity. This is important for diverse models such as the
She-L\'ev\^eque model~\cite{she-leveque:1994}, where physical
reasoning is based on the one hand on the energy dissipation rate and
the RSH and on the other hand on the dimensionality of the coherent
structures of vorticity. This question is also closely related to the
issue of different scaling laws for longitudinal and transverse
structure functions as we will explain in the next section.

\begin{figure}[h]
  \begin{center}
    \includegraphics[width=0.8\columnwidth]{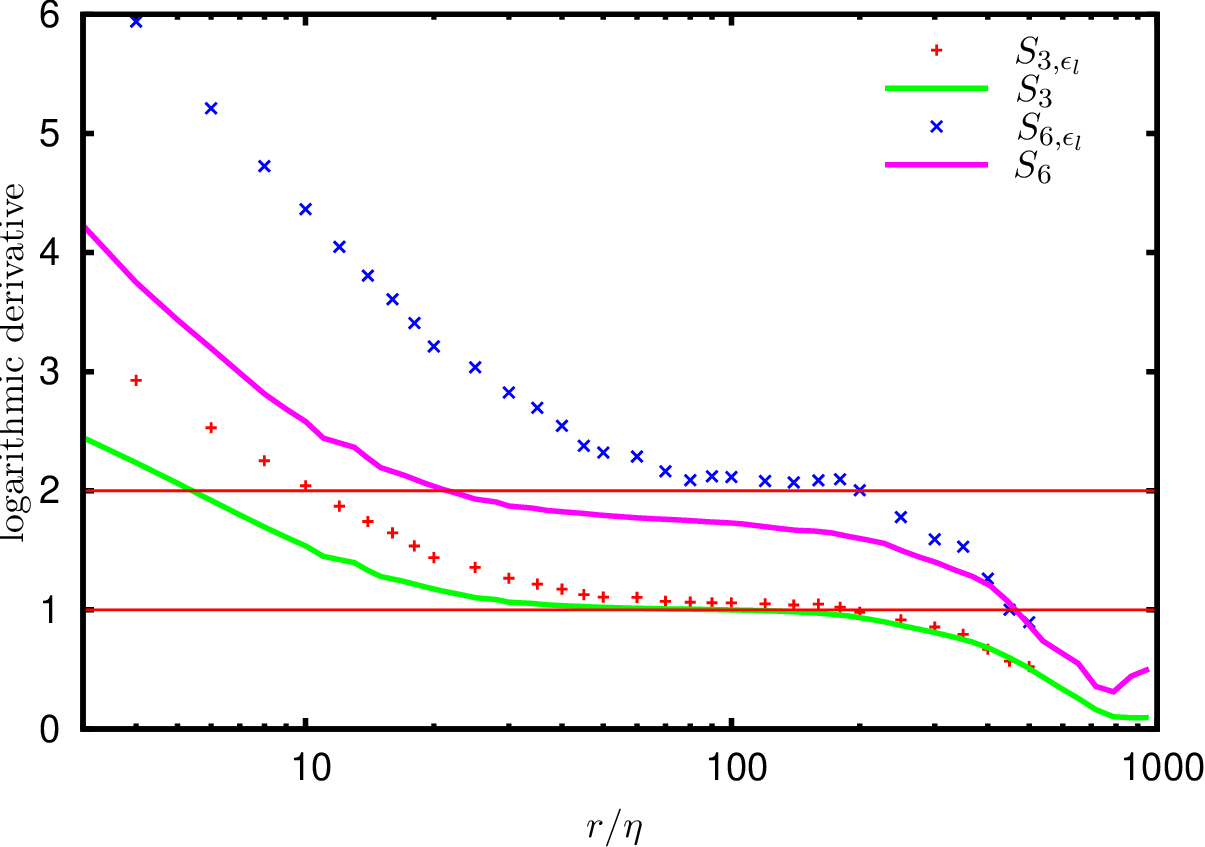} 
  \end{center}
  \caption{\label{fig:strucfunc_vgl} Logarithmic derivative of the
    conditioned and unconditioned longitudinal velocity structure
    function of order $p=3$ and $p=6$, lines indicate the
    K41-prediction}
\end{figure}

We conclude this section on longitudinal increments by an examination
of the corresponding conditioned structure functions
$S_{p,\epsilon_l}$. From the scale-invariant PDFs in
Fig.~\ref{fig:AllFlat} we expect them to follow the linear K41-scaling
law $p/3$ within the inertial range. Indeed, as shown in
Fig.~\ref{fig:strucfunc_vgl}, the conditioned structure functions
follow Kolmogorov's prediction while the unconditioned higher-order
functions exhibit lower plateaus expressed by a non-zero $\mu_p$
in~(\ref{eq:correction}).

\subsection{Transverse increments} 

In analogy to the RSH for the longitudinal velocity increments Chen et
al.~\cite{CHE97} proposed a refined self-similarity hypothesis for the
transverse velocity increments (RSHT). This relation of the
scale-averaged square of vorticity and the transverse velocity
increments reads
\begin{equation}
  \delta^{\perp}_lu=\beta_2(\Omega_l l)^{\frac{1}{3}},  
\end{equation}
where $\beta_2$ is a statistical variable independent of $l$ and
$\Omega_l$, given by (\ref{equ:vorticityInt}).

\begin{figure}[h]
  \begin{center}
    \includegraphics[width=0.8\columnwidth]{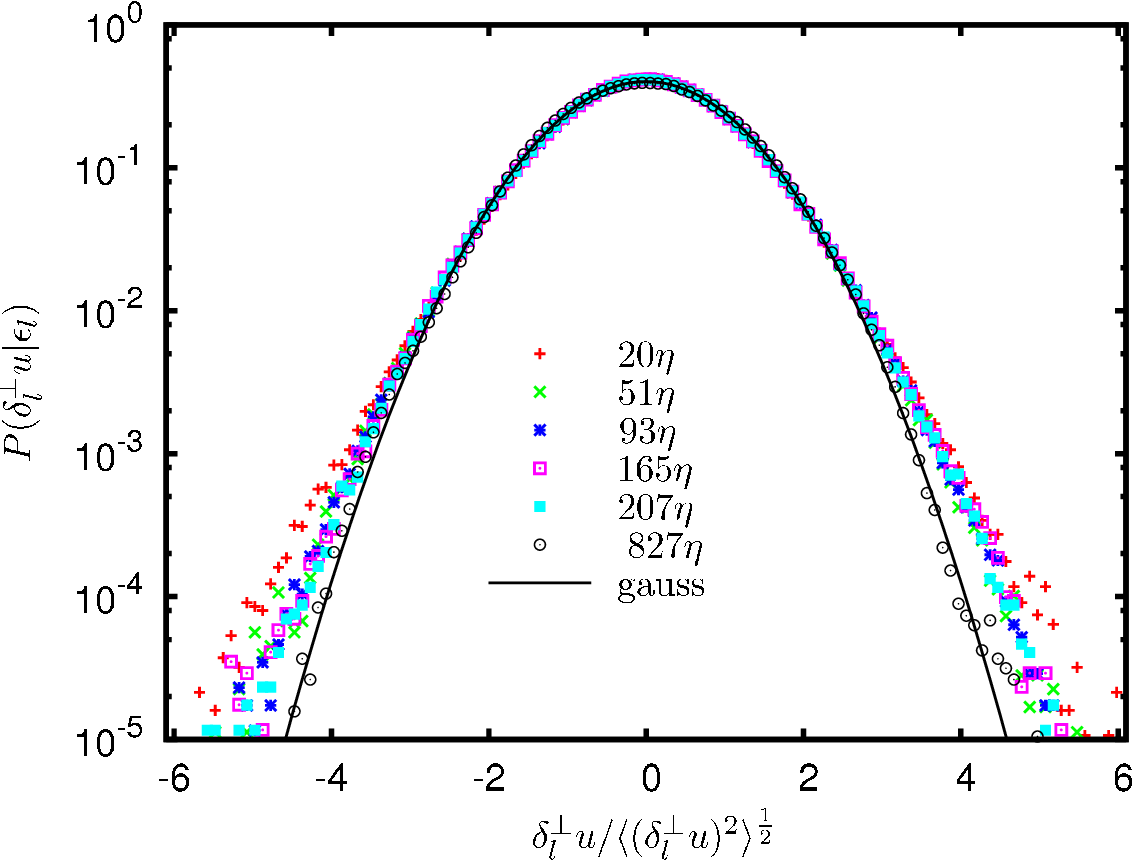} 
  \end{center}
  \caption{\label{fig:pdfsTrans} Conditioned PDFs $P(\delta^{\perp}_l
    u |\epsilon_l)$ for different separations $l$ in comparison to a
    Gaussian distribution, normalized to unit variance}
\end{figure}

\begin{figure}[h]
  \begin{center}
    \includegraphics[width=0.8\columnwidth]{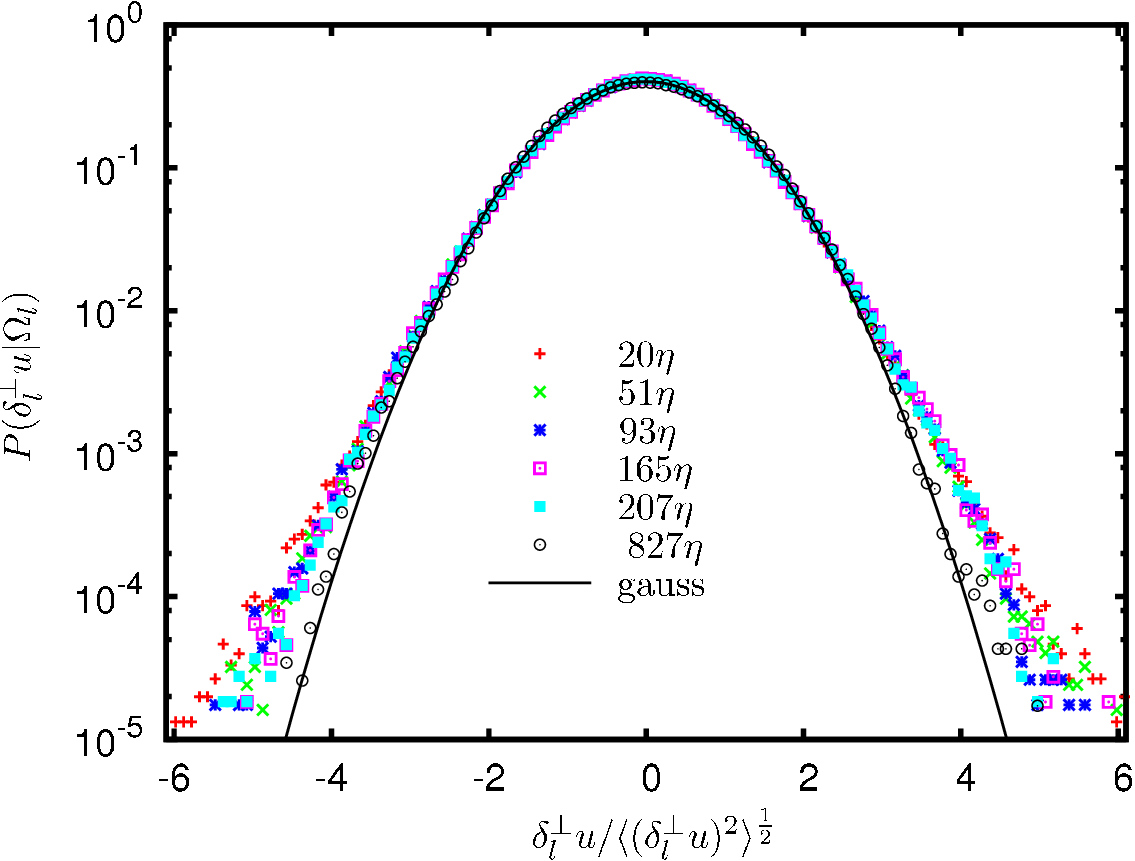} 
  \end{center}
  \caption{\label{fig:pdfsTransV} Conditioned PDFs $P(\delta^{\perp}_l
    u |\Omega_l)$ for different separations $l$ in comparison to a
    Gaussian distribution, normalized to unit variance}
\end{figure}

\begin{figure}[h]
  \begin{center}
    \includegraphics[width=0.8\columnwidth]{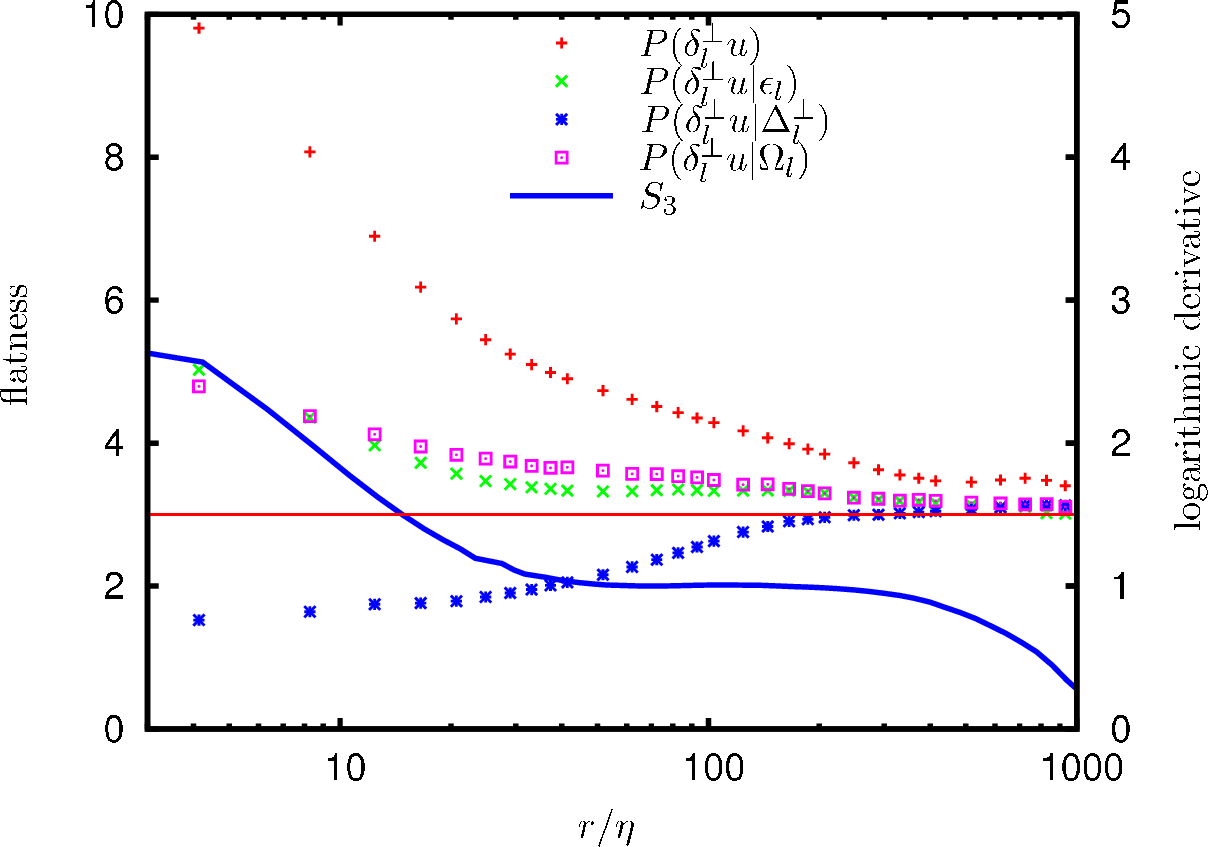} 
  \end{center}
  \caption{\label{fig:AllFlatT} Flatness factors of the conditioned
    PDFs $P(\delta^{\perp}_lu|\Omega_l)$,
    $P(\delta^{\perp}_lu|\epsilon_l)$,
    $P(\delta^{\perp}_lu|\Delta^\perp_l)$ and unconditioned
    $P(\delta^{\perp}_lu)$, including the third order transverse
    structure function $S^{\perp}_3$}
\end{figure}

\begin{figure}[h]
  \begin{center}
    \includegraphics[width=0.8\columnwidth]{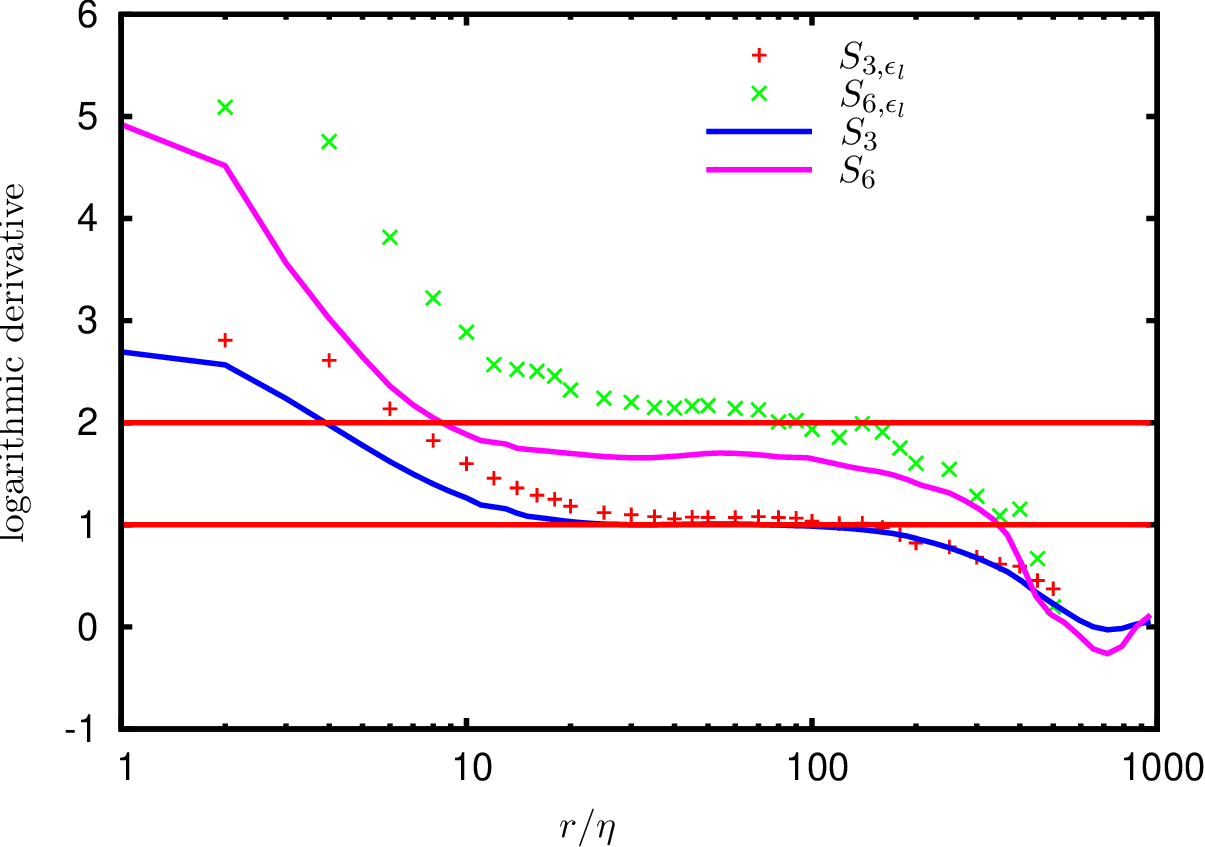} 
  \end{center}
  \caption{\label{fig:strucFuncT} Logarithmic derivative of the
    conditioned and unconditioned transverse velocity structure
    function of order $p=3$ and $p=6$, lines indicate the Kolmogorov
    prediction}
\end{figure}

Following Chen et al. it is reasonable to look in the transverse case
at the statistics of velocity increments conditioned to $\Omega_l$,
namely the PDFs $P(\delta^{\perp}_l u|\Omega_l)$. They are
scale-invariant and only slightly flatter than Gaussian PDFs (see
Fig.~\ref{fig:pdfsTrans}). As in the case of longitudinal increments
one can ask how other conditions such as the energy dissipation rate
perform compared to $\epsilon_l$. The PDFs conditioned on $\epsilon_l$
shown in Fig.~\ref{fig:pdfsTransV} are indistinguishable from the PDFs
conditioned on the vorticity $\Omega_l$. From this point of view it is
impossible to conclude whether $\epsilon_l$ or $\Omega_l$ is the
better condition for the transverse fluctuations.
In order to make a more precise statement on the difference between
these two conditions it is helpful to look at the flatness in
Fig.~\ref{fig:AllFlatT}.
From this we conclude that conditioning to $\epsilon_l$ or $\Omega_l$
yields quasi-identical results and surprisingly $P(\delta^{\perp}_l u
|\epsilon_l)$ are even slightly more Gaussian than $P(\delta^{\perp}_l
u |\Omega_l)$. The averaged transverse gradient
\begin{equation}
  \Delta^{\perp}_l=\frac{1}{l} \int^l_0ds\, \nu |\hat{\bm{l}}\times\nabla
  \bm{u}(\bm{x}+s\,\hat{\bm{l}})|^2.
  \label{equ:gradientPerp}
\end{equation} 
reduces the flatness less than the two other conditions.
The transverse structure functions are shown in
Fig.~\ref{fig:strucFuncT}. As expected, we find that the conditioned
ones follow the K41 prediction with the inertial range of scale while
the high-order unconditioned functions have significantly lower
plateaus.

\section{Lagrangian conditional statistics}
\label{sec:LAGRANGIAN}
After having computed statistics in the Eulerian framework we now
consider velocity increments (\ref{eq:lagInc}) in the Lagrangian frame
of reference. The Lagrangian analog to the RSH might be labeled the
Lagrangian refined self-similarity hypothesis (LRSH) and reads
\begin{equation}
  \label{eq:lagRSH}
  \delta_\tau v_i = \beta_L(\tau \epsilon_\tau)^{1/2},
\end{equation}
where the local energy dissipation rate (\ref{eq:localEnergyDiss}) is
averaged along a particle trajectory according to
\begin{equation}
  \label{eq:LagEpsilon}
  \epsilon_\tau=\frac{1}{\tau}\int_0^\tau \epsilon(\bm{X}(\bm{x}_0,t))dt.
\end{equation}
However, one can question whether $\epsilon_\tau$ is the correct
quantity appearing in (\ref{eq:lagRSH}). Benzi et al.~\cite{Benzi2009}
examined this relation by means of the assumption of extended
self-similarity and found that $\epsilon_\tau$ rather than the
averaged square of vorticity
\begin{equation*}
  \Omega_\tau = \frac{1}{\tau}\int_0^\tau |\bm{\omega}(\bm{X}(\bm{x}_0,t))|dt.
\end{equation*}
is the correct quantity in (\ref{eq:lagRSH}). Yu et al.~\cite{Yu2010}
conditioned the velocity increments on a spatially averaged energy
dissipation rate at one foot-point of the increments.  

In this work we stick to trajectory-averaged conditions and propose
yet another one for Lagrangian increment statistics. In order to
motivate this on dimensional grounds we recall that Eulerian
increments $(u_i(l\,\bm{e}_j)-u_i(0))/l$ tend to spatial derivatives
$\partial_j u_i$ of the velocity field in the limit $l\rightarrow
0$. Those derivatives appear in the local energy dissipation rate
(\ref{eq:localEnergyDiss}). Instead, Lagrangian increments
$(u_i(\tau)-u_i(0))/\tau$ tend to the fluid-particle acceleration in
the limit $\tau \rightarrow 0$ which involve a term $u_j\partial_j
u_i$.  We therefore propose to replace (\ref{eq:LagEpsilon}) by
\begin{equation}
\epsilon^L_\tau = \frac{1}{2} \int dt \sum_{i,j}[u_j\,\partial_j
  u_i + u_i\,\partial_i u_j]^2
\label{equ:lag-dissipation}
\end{equation}
in the LRSH (\ref{eq:lagRSH}).

The calculation of $\epsilon_\tau$, $\Omega_\tau$, and
$\epsilon^L_\tau$ for a given time lag $\tau$ is done by averaging the
local quantities over all stored points along the particle
trajectory. We achieved converged statistics by taking the average
over 10 Million particles and several large-eddy turn-over times.

\begin{figure}[h]
  \begin{center}
    \includegraphics[width=0.8\columnwidth]{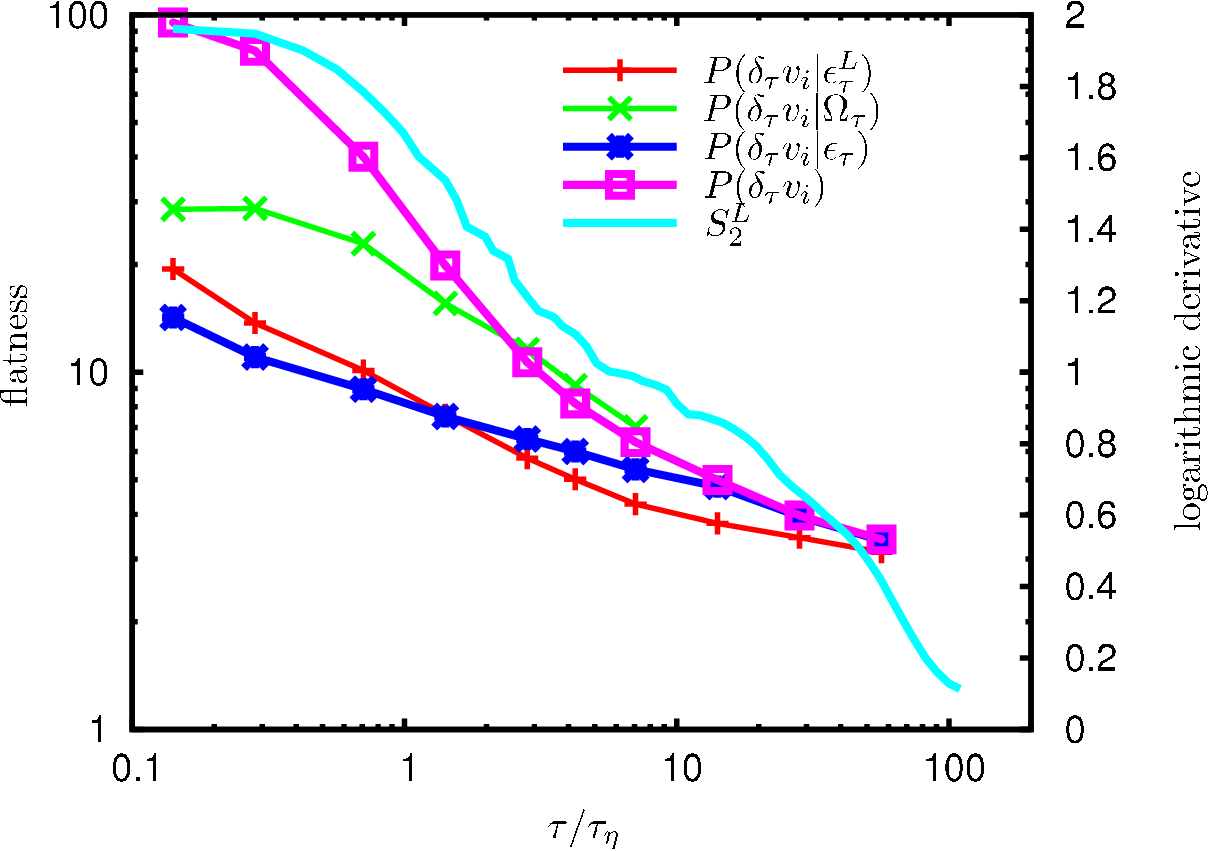} 
  \end{center}
  \caption{\label{fig:lagFlat} Flatness factors of the conditioned
    velocity increment PDFs $P(\delta_\tau v_i|\epsilon_\tau)$,
    $P(\delta_\tau v_i|\epsilon^L_\tau)$ and $P(\delta_\tau
    v_i|\Omega_\tau)$ as well as of the unconditioned PDF
    $P(\delta_\tau v_i)$ together with the logarithmic derivative of
    $S^L_2$}
\end{figure}

In Fig.~\ref{fig:lagFlat} we compare the flatness of velocity
increment PDFs conditioned on $\epsilon_\tau$, $\Omega_\tau$, and
$\epsilon^L_\tau$. We added the logarithmic derivative of the second
order Lagrangian structure function $S^L_2(l) = \langle (\delta_\tau
v_i)^2\rangle$ in order to clarify three different ranges of scales:
The dissipative scales up to $\tau \approx 1$, the inertial ones $1 <
\tau < 60$, followed by the large scales. If we restrict our attention
to the inertial range we observe that the flatness is most efficiently
reduced by $\epsilon^L_\tau$. Also the trajectory integrated energy
dissipation rate $\epsilon_\tau$ diminishes significantly the flatness
while the integrated vorticity $\Omega_\tau$ has a negligible
effect. This indicates that $\epsilon^L_\tau$ might be a more
appropriate condition than $\epsilon_\tau$.

\begin{figure}[h]
  \begin{center}
    \includegraphics[width=0.8\columnwidth]{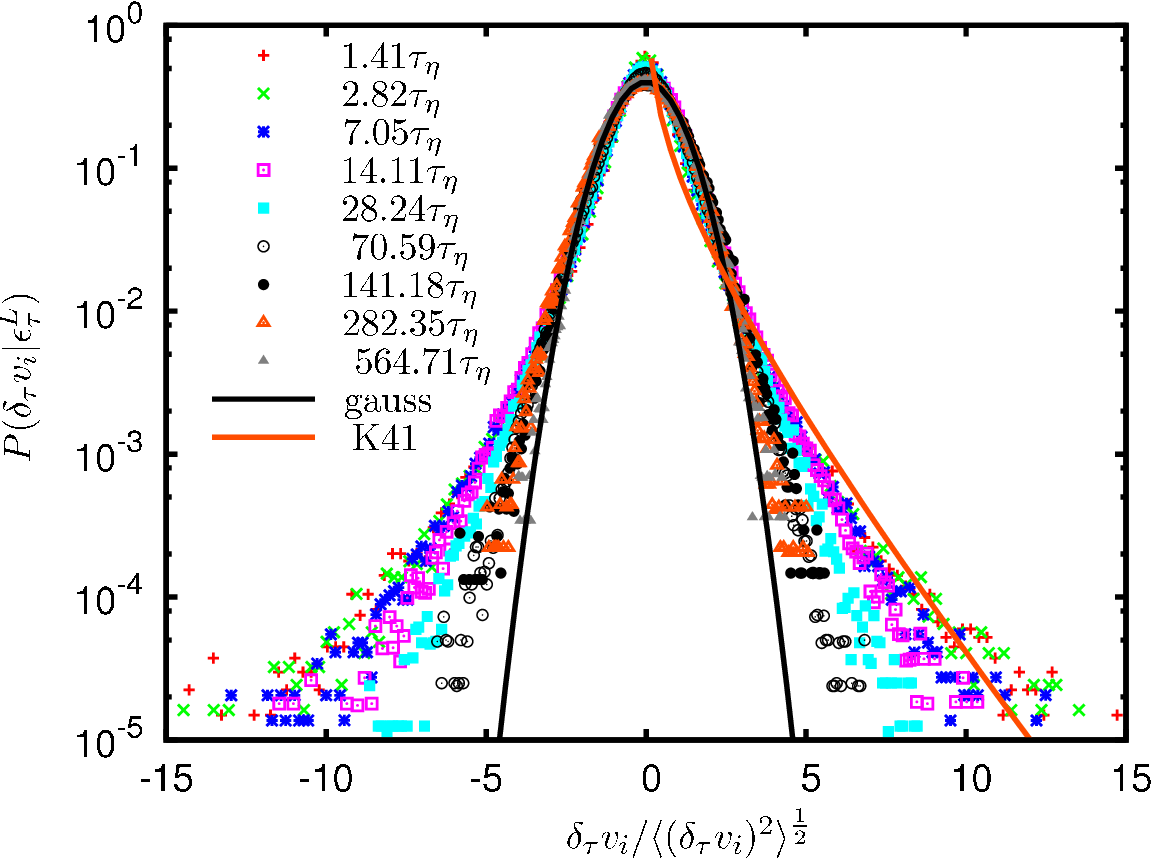} 
  \end{center}
  \caption{\label{fig:pdfsLAG} Conditioned PDFs $P(\delta_\tau
    v_i|\epsilon_\tau)$ for different time lags $\tau$ in comparison
    to a Gaussian distribution and to the K41-prediction for the PDF
    of acceleration, normalized to unit variance}
\end{figure}

The corresponding conditioned increment PDFs are labeled
$P(\delta_\tau v_i|\epsilon_\tau)$ and show in
Fig.~\ref{fig:pdfsLAG}. The PDF corresponding to the shortest time-lag
considered is reasonably well described by the K41-acceleration
PDF~\cite{biferale:2004b}
\begin{equation}
  P(a)=(a/b)^{-5/9}\exp[-0.5\, (a/b)^{8/9}]/c
\end{equation}
normalized to unit-variance with $a=0.48$ and $b=2.72$.  This PDF is
the Lagrangian analogon to a Gaussian distribution for Eulerian
velocity gradients.

It is important to note that contrarily to the results in Eulerian
setup the conditioned Lagrangian PDFs $P(\delta_\tau
v_i|\epsilon^L_\tau)$ (see Fig.~\ref{fig:pdfsLAG}) are still
scale-dependent. One notes a transition from stretched tails
(K41-prediction) for short time-lags to Gaussian PDFs (uncorrelated
statistics) for time lags of the order of the integral time
scale. This implies that Lagrangian increment statistics is
'naturally' scale dependent.

\begin{figure}[h]
  \begin{center}
    \includegraphics[width=0.8\columnwidth]{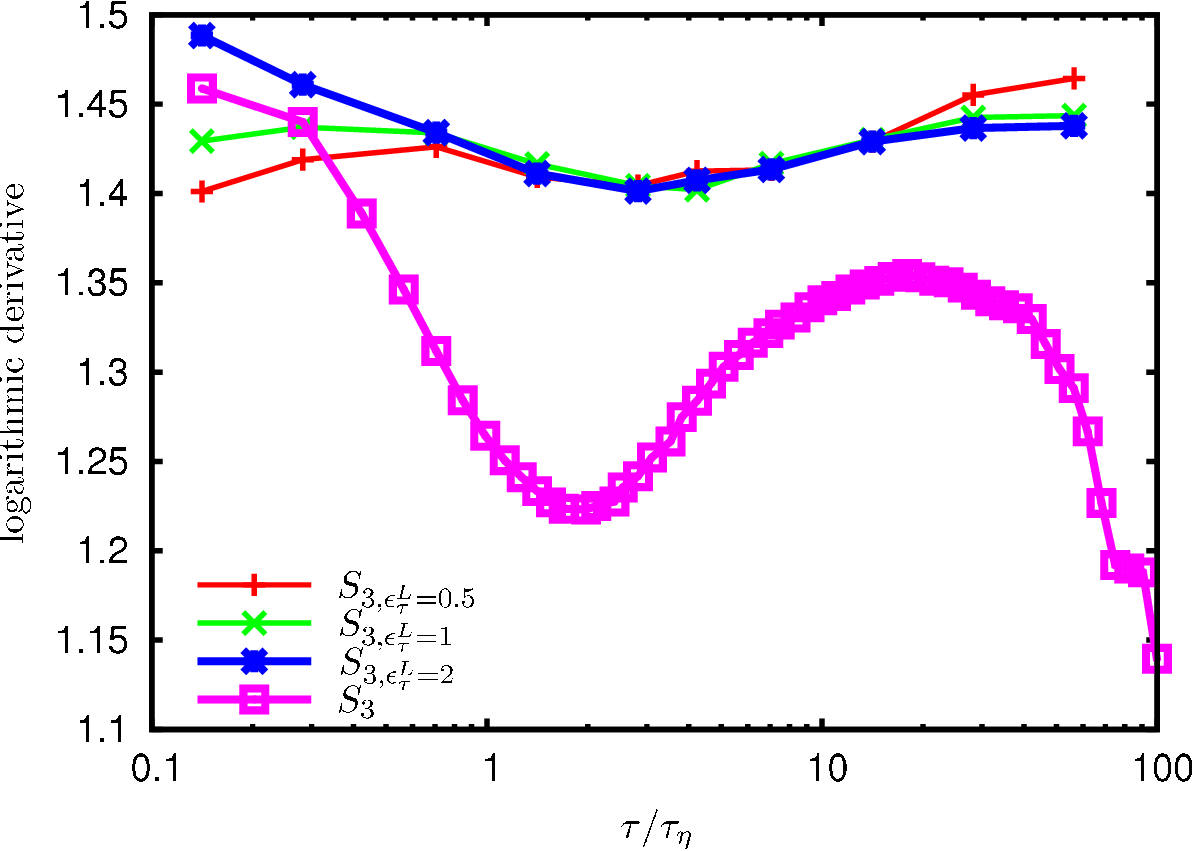} 
  \end{center}
  \caption{\label{fig:LagStrucFunc} Logarithmic derivatives of
    relative Lagrangian velocity structure
    function. $\epsilon^L_\tau=1$ corresponds to the one with the most
    statistics}
\end{figure}

As can be see from the unconditioned structure function in
Fig.~\ref{fig:lagFlat}, Lagrangian structure functions do not show a
clear scaling law at today accessible Reynolds numbers. We therefore
refer to relative structure functions $S_p(S_2)$. In the computation
of the conditioned structure functions we fixed one $\epsilon^L_\tau$
for all increments $\tau$. In Fig.~\ref{fig:LagStrucFunc} their
logarithmic derivatives are shown which clearly change under the
condition $\epsilon^L_\tau$. There are two major differences between
the conditioned and unconditioned functions. The first concerns
intermittency: The conditioned functions have larger values than the
unconditioned ones. We observe a value of approximately. 1.43 which is
close to the K41 prediction of 1.5. This implies that intermittency is
significantly reduced on subsets $\Omega_{\epsilon^L_\tau}$. A second
feature of Lagrangian increment statistics is the so called bottleneck
around a few $\tau_\eta$. It has been attributed to the characteristic
trajectories (spirals) of tracers in the vicinity of coherent vortex
filaments. This bottleneck in the local slop is absent once velocity
increments are conditioned (see again Fig.~\ref{fig:LagStrucFunc}),
which means that their scaling range is enlarged. Its origin is
supposed to be in the coexistence of two different power-laws. The
first related to dissipative effects and the second to inertial range
physics\cite{ictr:2008}. An insufficient separation of dissipative and
inertial scales might lead to the observed dip in the local slope of
structure functions. Interestingly, this bottleneck is negligible in
the case of conditioned structure functions. This implies that it is
due to a mixture of statistics from different subset
$\Omega_{\epsilon^L_\tau}$.

\section{Conclusion}
\label{sec:CONLUSION}
This work investigates the statistics of Eulerian and Lagrangian
velocity increments when conditioned to different scale-averaged
quantities such as the energy dissipation rate, the square of
vorticity or the velocity gradient. In the case of Lagrangian
increments we propose a novel condition dimensionally related to the
acceleration of fluid elements.

Considering Eulerian statistics we find that longitudinal as well as
transverse increment PDFs are Gaussian shaped with flatness factors
close to three when conditioned to the scale-averaged energy
dissipation rate. The averaged vorticity produces slightly flatter
tails while the longitudinal and transverse velocity gradient perform
significantly worse. Therefore, there is no preferential link of
transverse increments and vorticity as of longitudinal increments and
energy dissipation rate which is important for models of
intermittency. Conditional structure functions show clear K41-scaling
within the inertial range of scales.

Considering Lagrangian statistics we investigated velocity increments
conditioned to trajectory-averaged quantities such as the energy
dissipation rate, the vorticity and a novel condition. The latter is
motivated by dimensional arguments. Conditioning to the dissipation
rate and to the novel condition yields flatnesses of the increment PDFs
much smaller than without conditioning. More precisely, the
conditioned PDF of the shortest increment considered agrees reasonably
well with the K41-prediction for the PDF of acceleration. Within the
inertial range of scales the flatnesses of PDFs under the novel
condition are even smaller than the flatness of PDFs conditioned on
the averaged energy dissipation.

Conditioned and unconditioned Lagrangian structure functions differ
significantly. First, conditioning yields quasi-K41 scaling
exponents. Secondly, the characteristics bottleneck of the
unconditioned functions at the onset of the inertial range disappears
once conditioned.
\newline \\
\noindent {\sc Acknowledgments.} This study benefited from fruitful
discussions with A. Naert and J. Bec. Access to the IBM BlueGene/P
computer JUGENE at the FZ J\"ulich was made available through the
'XXL-project' of HBO28 and partly through project HBO36. This work
benefited from support through DFG-FOR1048

\bibliographystyle{ieeetr}

\begin{thebibliography}{10}

\bibitem{frisch:1995}
U.~Frisch, {\em Turbulence} (Cambridge, Cambridge University Press, 1995).

\bibitem{tsinober:2009}
A.~Tsinober, {\em An Informal Conceptual Introduction to Turbulence: Second
  Edition of An Informal Introduction to Turbulence}. (Springer, Netherlands, 2009).

\bibitem{K41}
A.~N. Kolmogorov {\em C. R. Acad. Sci. URSS}, {\bf 32}, 19 (1941).

\bibitem{Dhruva2000}
B.~Dhruva, Y.~Tsuji, and K.~Sreenivasan, ``{Transverse structure functions in
  high-Reynolds-number turbulence},'' {\em Physical Review E}, {\bf 56},
  4928--4930 (2000).

\bibitem{Zhou2005a}
T.~Zhou, Z.~Hao, L.~P. Chua, and S.~C.~M. Yu, ``{Scaling of longitudinal and
  transverse velocity increments in a cylinder wake},'' {\em Physical Review
  E}, {\bf 71}, 066307 (2005).

\bibitem{Grossmann1997}
S.~Grossmann, L.~Detlef, and A.~Reeh, ``{Different intermittency for
  longitudinal and transversal turbulent fluctuations},'' {\em Physics of
  Fluids}, {\bf 9}, 3817 (1997).

\bibitem{Germaschewski1999}
K.~Germaschewski and R.~Grauer, ``{Longitudinal and transversal structure
  functions in two-dimensional electron magnetohydrodynamic flows},'' {\em
  Physics of Plasmas}, {\bf 6}, 3788 (1999).

\bibitem{grauer-homann-etal:2010}
R.~Grauer, H.~Homann, and J.-F. Pinton, ``On longitudinal and transverse
  structure functions in high reynolds-number turbulence,'' {\em in
  preparation}.

\bibitem{K62}
A.~N. Kolmogorov {\em J. Fluid Mech.}, {\bf 13}, 82 (1962).

\bibitem{GAG94}
Y.~Gagne, M.~Marchand, and B.~Castaing, ``Conditional velocity pdf in 3-d
  turbulence,'' {\em J. Phys. II France}, {\bf 4}, 1--8 (1994).

\bibitem{ott-mann:2000}
S.~Ott and J.~Mann, ``An experimental investigation of the relative diffusion
  of particle pairs in three-dimensional turbulent flow,'' {\em Journal of
  Fluid Mechanics}, {\bf 422}, 207--223 (2000).

\bibitem{porta-bodenschatz-etal:2001}
A.~L. Porta, G.~A. Voth, A.~M. Crawford, J.~Alexander, and E.~Bodenschatz,
  ``Fluid particle accelerations in fully developed turbulence,'' {\em Nature},
  {\bf 409}, 1017--1019 (2001).

\bibitem{mordant:2001}
N.~Mordant, P.~Metz, O.~Michel, and J.~F. Pinton, ``Measurement of lagrangian
  velocity in fully developed turbulence,'' {\em Phys. Rev. Lett.}, {\bf 87},
  214501 (2001).

\bibitem{yeung-pope-etal:2006}
P.~K. Yeung and M.~S. Sawford, ``Reynolds number dependence of lagrangian
  statistics in large numerical simulations of isotropic turbulence,'' {\em J.
  of Turbulence}, {\bf 7}, 1--12 (2006).

\bibitem{Biferale2008}
L.~Biferale, E.~Bodenschatz, M.~Cencini, A.~Lanotte, N.~T. Ouellette,
  F.~Toschi, and H.~Xu, ``{Lagrangian structure functions in turbulence: A
  quantitative comparison between experiment and direct numerical
  simulation},'' {\em Phys. Fluids}, {\bf 20}, 065103 (2008).

\bibitem{biferale:2004b}
L.~Biferale, G.~Boffetta, A.~Celani, B.~J. Devenish, A.~Lanotte, and F.~Toschi,
  ``Multifractal statistics of lagrangian velocity and acceleration in
  turbulence,'' {\em Phys. Rev. Lett.}, {\bf 93}, 4502 (2004).

\bibitem{kamps-friedrich-grauer:2009}
O.~Kamps, R.~Friedrich, and R.~Grauer, ``An exact relation between eulerian and
  lagrangian velocity increment statistics,'' {\em Phys. Rev. E}, {\bf 79}, 066301 (2009).

\bibitem{homann-kamps-etal:2009}
H.~Homann, O.Kamps, R.~Friedrich, and R.~Grauer, ``Bridging from eulerian to
  lagrangian statistics in 3d hydro- and magnetohydrodynamic turbulent flows,''
  {\em New J. Phys.}, {\bf 11}, 073020 (2009).

\bibitem{NAE98}
A.~Naert, B.~Castaing, B.~Hébral, and J.~Peinke, ``Conditional statistics of
  velocity fluctuations in turbulence,'' {\em Physica D}, {\bf 113}, 73--78 (1998).

\bibitem{hou-li:2007}
T.~Y. Hou and R.~Li, ``Computing nearly singular solutions using
  pseudo-spectral methods,'' {\em J. Comp. Phys}, {\bf 226}, 379--397 (2007).

\bibitem{grafke-homann-etal:2007}
T.~Grafke, H.~Homann, J.~Dreher, and R.~Grauer, ``Numerical simulations of
  possible finite time singularities in the incompressible euler equations:
  comparison of numerical methods,'' {\em Physica D}, {\bf 237}, 1932--1936 (2008).

\bibitem{p3dfft}
``Parallel 3d fast fourier transforms (p3dfft).''
  http://www.sdsc.edu/us/resources/p3dfft.

\bibitem{shu-osher:1988}
C.~Shu and S.~Osher, ``Efficient implementation of essentially non-oscillatory
  shock-capturing schemes,'' {\em J. Comput. Phys.}, {\bf 77}, 439--471 (1988).

\bibitem{homann-dreher-etal:2007}
H.~Homann, J.~Dreher, and R.~Grauer, ``Impact of the floating-point precision
  and interpolation scheme on the results of dns of turbulence by
  pseudo-spectral codes,'' {\em Comput. Phys. Comm.}, {\bf 177}, 560--565 (2007).

\bibitem{she-leveque:1994}
Z.-S. She and E.~L\'ev\^eque, ``Universal scaling laws in fully developed
  turbulence,'' {\em Phys. Rev. Lett.}, {\bf 72}, 336--339 (1994).

\bibitem{CHE97}
S.~Chen, K.~Sreenivasan, M.~Nelkin, and N.~Cao, ``Refined similarity hypothesis
  for transverse structure functions in fluid turbulence,'' {\em Phys. Rev.
  Lett.}, {\bf 79}, 2253 (1997).

\bibitem{Benzi2009}
R.~Benzi, L.~Biferale, E.~Calzavarini, D.~Lohse, and F.~Toschi,
  ``{Velocity-gradient statistics along particle trajectories in turbulent
  flows: The refined similarity hypothesis in the Lagrangian frame},'' {\em
  Physical Review E}, {\bf 80}, 066318 (2009).

\bibitem{Yu2010}
H.~Yu and C.~Meneveau, ``{Lagrangian Refined Kolmogorov Similarity Hypothesis
  for Gradient Time Evolution and Correlation in Turbulent Flows},'' {\em
  Physical Review Letters}, {\bf 104}, 084502 (2010).

\bibitem{ictr:2008}
ICTR, A.~Arneodo, R.~Benzi, J.~Berg, L.~Biferale, E.~Bodenschatz, A.~Busse,
  E.~Calzavarini, B.~Castaing, M.~Cencini, L.~Chevillard, R.~Fisher, R.~Grauer,
  H.~Homann, D.~Lamb, A.~Lanotte, E.~Leveque, B.~Luethi, J.~Mann, N.~Mordant,
  W.~Mueller, S.~Ott, N.~Ouellette, J.~Pinton, S.~Pope, S.~Roux, F.~Toschi,
  H.~Xu, and P.~Young, ``Universal intermittent properties of particle
  trajectories in highly turbulent flows,'' {\em Phys. Rev. Lett.}, {\bf 100},
  254504 (2008).

\end{thebibliography}

\end{document}